\documentclass{article}
\usepackage{amsmath}
\newcommand{\sub}[1]{_{\scriptscriptstyle{#1}}}
\renewcommand{\sup}[1]{^{\scriptscriptstyle{#1}}}
\newcommand{\pare}[1]{\left(#1\right)}
\begin{document}
\begin{flushright}
Preprint \qquad \qquad hep-th/9903016
\end{flushright}
\vskip 1truein
\begin{center}
{\huge ON PATH DEPENDENT STATE SPACE\\
 FOR THE PROCA FIELD}
\vskip 5pt
{\large ROLANDO GAITAN}{\footnote {email: rgaitan@fisica.ciens.ucv.ve} } \\
{\it Grupo de F\'{\i}sica Te\'orica, Departamento de F\'{\i}sica,\\
Facultad de Ciencias y Tecnolog\'{\i}a, Universidad de Carabobo,\\
Valencia, Venezuela }\\
\end{center}
\vskip .5truein
\begin{abstract}
A gauge formulation for the Proca model quantum theory in an open path
functional space representation is revisited. The path dependent vacuum
state is obtained. Starting from this one, other excited states can be
obtained too. Additionally, the functional integration measure needed to
define an internal product in the state space is constructed.
\end{abstract}
In the reference \cite{CGL} authors have presented the Proca gauge theory in
a ``Geometric'' form. They use an open path representation \cite{CGL,FG,GL}
 ($P$-representation) in wich the general quantum states and operators
depend and act on path functionals, respectively. In this letter, as a
contribution to the understanding of this formulation we discuss the
explicit construction of the open path state space for Proca gauge theory as
a geometrical extension from standard Quantum Field Theory. The letter is
organized as follow. We present a self contained brief review on Proca gauge
Theory in a Geometric Representation \cite{CGL}. Then, the vacuum state is
presented and, from its formal normalization the funtional integration
measure is deduced. Finally, a recurrence relation for the excited states is
given , showing as an example the first excited state and its energy.

The brief review begins here. Let $L$ be the Lagrangian for the Proca gauge
theory
\begin{equation}
L=\left\langle -\frac 14 F\sub{\mu\nu}F\sup{\mu\nu}+\frac 12M\sup
2\pare{A\sub\mu+\partial\sub\mu\omega}\pare{A\sup\mu+\partial\sup\mu\omega}
\right\rangle,\label{eq1}
\end{equation} 
where the symbol $\left\langle\cdots\right\rangle$ means $\int d\sup
3x(\cdots)$,
$F\sub{\mu\nu}=\partial\sub\mu A\sub\nu-\partial\sub\nu A\sub\mu$ is the
Maxwell field and $\omega$ is an scalar field (i.e., St\"uckelberg Scalar).
Eq. \eqref{eq1} is gauge invariant under the transformations $\delta
A\sub\mu=\partial\sub\mu\Lambda$ and $\delta\omega=-\Lambda$. Subsequently,
form Dirac's quatization, First Class primary and secondary constraints
arise; namely $\pi\sup 0\sub A\approx 0$ and $\partial\sub k\pi\sup 0\sub
A+\pi\sub\omega\approx 0$, where the $\pi$'s are the canonical conjugates 
momenta of Proca and scalar fields, respectively. The Hamiltonian can then 
be obtained in the standard way
\begin{equation}
H=\left\langle\frac 12{\pare{\pi\sup i\sub A}}\sup 2+\frac 1{2M\sup
2}{\pare{\pi\sub\omega}}\sup 2+\frac 14{\pare{F\sub{ij}}}\sup 2+\frac 12
M\sup 2{\pare{A\sub i+\partial\sub i\omega}}\sup 2\right\rangle.\label{eq2}
\end{equation}

The quantization of the theory starts with the promotion from fields to
operators acting on the Hilbert space physical sector in the
$P$-representation $\pare{H\sub p}$ and the Poisson algebra to a commutator
one. The sector $H\sub p$ is defined trough $P$-functionals prescribing
$\Psi(P.Q)=f(P)\Psi(Q)$ (here $P$ and $Q$ are open path colections) where
\begin{equation}
f(P\sub x\sup y)=e\sup{-i\omega(x)}\exp{\pare{i\int\sub Pdt\sup kA\sub k(t)}}
e\sup{i\omega(y)}\label{eq3}
\end{equation}
is the gauge invariant ``Wilson Path'' (following the ``Wilson Loop'' idea)
operator asociated to the $3$-spatial path $P\sub x\sup y$, wich starts at
$x$ and ends at $y$.

The operator \eqref{eq3} guarantees that $H\sub p$ contitutes a
$P$-functional set of solutions of the First Class Constraints. At the same
time $\Psi(P)$ are eigenfunctions of $\pi\sub\omega$ and $\pi\sup k\sub A$
operator with eigenvalues $\rho(z,P)$ and $T\sup k(z,P)$, respectively. Let
us recall the ``end densities'' 
\begin{equation}
\rho(z,P)=\sum_{a=1}^N\rho(z,P_a)\quad\hbox{and}\quad\rho(z,P_a)=\delta\sup
3(z-y_a)-\delta\sup 3(z-x_a),
\end{equation}
and the distributional field of tangent vectors associated to the paths
\begin{equation}
T\sup k(z,P)=\sum_{a=1}^NT\sup
k(z,P_a)\quad\hbox{and}\quad T\sup k(z,P_a)=\int\sub{P_a}dt\sup k\delta\sup
3(t-z).
\end{equation}

Next, when Mandelstam \cite{M} and Loop \cite{DNGT} derivatives are taken
into account, the Hamiltonian \eqref{eq2} can be rewritten in a geometric 
form
\begin{equation}
H=\left\langle\frac 12{\pare{T\sup i(z,P)}}\sup 2+\frac 1{2M\sup 2}\rho\sup
2(z,P)-\frac 14{\pare{\Delta\sub{ij}(z)}}\sup 2-\frac 12 M\sup
2{\pare{\Delta\sub i(z)}}\sup 2\right\rangle.\label{eq5}
\end{equation}

Let us remember that given a $P$-functional $G(P)$ the Mandelstam derivative
can be defined as
\begin{equation}
\delta P\sup i\Delta\sub i(z)G(P)=G(\delta P.P)-G(P),
\end{equation}
where $\delta P$ is an infinitesimal open path. On the other hand the Loop
derivative is
\begin{equation}
\delta\sigma\sup{ij}\Delta\sub{ij}(z)G(P)=G(\delta C.P)-G(P),
\end{equation}
where $\delta C$ is an infinitesimal Loop with area $\delta\Sigma\sub
k=\frac 12\epsilon\sub{kij}\delta\sigma\sup{ij}$. This ends the review.

In order to find the explicit form of the state space one can introduce the
vacuum state as follows. The operator (\ref{eq5}) has a cuadratic 
tangent-dependient interaction part and also end densities dependence. On
the other hand, the kinematic term is a Laplacian-like. Then, the key to
find the Schr\"oedinger eigenstates consists in carrying out a narrow
analogy with the standard quantum mechanic harmonic oscillator. For this
system, the ground state should have a Gaussian functional form
\begin{equation}
\Psi\sub 0(P)=\exp{\pare{-S(P)}}.\label{eq8}
\end{equation} 
Demanding $S(P)$ to be an even $P$-functional, a more general possible form
is
\begin{equation}
S(P)=\int\sub Pdy\sup i\int\sub P{dy'}\sup j\xi\sub{ij}(y-y'),\label{eq9}
\end{equation}
where the propagator $\xi\sub{ij}(y-y')$ is an explicit invariant under path
translations and satisfies the following properties
\begin{equation}
\hbox{symetry:
}\xi\sub{ij}(x)=\xi\sub{ji}(-x)\qquad\hbox{and}\qquad\hbox{reality:
}\xi\sub{ij}(x)=\xi\sub{ij}^*(x). 
\end{equation}

Considering $E\sub 0$ the ground state energy eigenvalue, then the equation
$H\Psi\sub 0(P)=E\sub 0\Psi\sub 0(P)$ gives the following relations
\begin{eqnarray}
&&\int\sub Pdy\sup i\int\sub P{dy'}\sup j\bigg\langle\frac{\delta\sup
3(z-y)\delta\sup 3(z-y')}2+\frac 1{2M\sup 2}\frac{\partial\delta\sup
3(z-y)}{\partial y\sup i}\frac{\delta\sup 3(z-y')}{\partial {y'}\sup
j}\nonumber\\
&&-\pare{\xi\sub{il,k}(y-z)-\xi\sub{ik,l}(y-z)}\pare{\xi\sub{jl,k}
(y'-z)-\xi\sub{jk,l}(y'-z)}-2M\sup 2\xi\sub{ik}(y-z)\xi\sub{jk}(y'-z)\bigg\rangle=0
\label{eq11}
\end{eqnarray}
and
\begin{equation}
\left\langle M\sup 2\xi\sub{kk}(0)-\xi\sub{kk,ll}(0)+\xi\sub{kl,kl}(0)\right
\rangle=E\sub
0,\label{eq12}
\end{equation}
where the propagator and energy are formally manipulated. Anyway equations
(\ref{eq11}) y (\ref{eq12}) gives the formal solution for the vacuum state.
To obtain this solution we transform to the Fourier Space, defining the
transformed propagator $A\sub{ij}$ by means of
\begin{equation}
\xi\sub{ij}(x)=\int\frac{d\sup 3p}{(2\pi)\sup 3}A\sub{ij}(p)e\sup{-ip\cdot
x}.\label{eq13}
\end{equation}
Substitution of eq. (\ref{eq13}) in (\ref{eq11}) leads to
\begin{equation}
\frac{\delta\sub{ij}}2+\frac 1{2M\sup 2}p\sub ip\sub
j-\pare{A\sub{il}(p)p\sub k-A\sub{ik}(p)p\sub l}\pare{A\sub{lj}(p)p\sub
k-A\sub{kj}(p)p\sub l}-2M\sup 2A\sub{ik}(p)A\sub{kj}(p)=0.\label{eq14}
\end{equation}
Here, the $A\sub{ij}(p)$ have some symetry properties coming from the
original propagator; namely $A\sub{ij}(p)=A\sub{ij}(-p)$ and
$A\sub{ij}(p)=A\sub{ij}^*(-p)$. Taking this into account, we choose the
ansatz
\begin{equation}
A\sub{ij}(p)=a(p\sup 2)\delta\sub{ij}+b(p\sup 2)p\sub ip\sub j,\label{eq15}
\end{equation}
where $a(p\sup 2)$ and $b(p\sup 2)$ are even real functions of momentum $p$.
These functions can be easily obtained if one substitutes eq. (\ref{eq15})
in (\ref{eq14}). This gives the following double solutions for $a$ and $b$
\begin{equation}
a\sub{\pm}=\pm\frac 1{2w(p)}\quad\hbox{and}\quad b\sub{\pm}=-\frac
a{p\sup2}\pm\frac{w(p)}{2M\sup 2p\sup 2},
\end{equation}
with $w(p)=\sqrt{p\sup 2+M\sup 2}$. However, if we require the Gaussian form
for the ground state, only the particular choice $a\sub{+}$ and $b\sub{-}$
solutions is allowed. Considering this combination, from (\ref{eq12}) the
vacuum state energy is obtained
\begin{equation}
E\sub 0=\left\langle\int\frac{d\sup 3p}{(2\pi)\sup 3}\pare{A\sub{kk}\pare{p\sup
2+M\sup 2}-A\sub{kl}p\sub kp\sub l}\right\rangle=\left\langle\frac 32\int\frac{d\sup
3p}{(2\pi)\sup 3}w(p)\right\rangle.
\end{equation}
It is noted that the $\frac 32$ factor, representing the three Proca field
degrees of freedom is recovered, as well as the Klein-Gordon dispersion
relation too. For this solution, the functional $S(P)$ yields
\begin{equation}
S(P)=\int \frac{d\sup 3p}{(2\pi)\sup 3}\int\sub P dy\sup i\int\sub P {dy'}\sup
j\frac{e\sup{-ip\cdot(y-y')}}{2w(p)}\pare{\delta\sub{ij}+\frac{p\sub i p\sub
j}{M\sup 2}}.\label{eq18}
\end{equation}
from this , the vacuum state is defined up to a conventional regularization
factor (i.e., cut-off, etc).

Additionally, eq. \eqref{eq18} throws light on the question of
$P$-functional space integration measure. This question is fundamental if
one wants to define an internal product in the state space. If we consider
the polarizations vector $\varepsilon\sup{(\lambda)}\sub\mu(p)$ ($\lambda$ 
runs from $1$ to $3$ and labels the polarization states) whose completeness
relations are
\begin{equation}
\varepsilon\sup{(\lambda)}\sub\mu(p)\varepsilon\sup{(\lambda)}\sub\nu(p)=
-\eta\sub{\mu\nu}+\frac{p\sub\mu p\sub\nu}{M\sup 2}
\end{equation}
with the Minkowski metric signature $\eta=diag(+---)$, one can define a
functional on $3$-spatial paths and momentum space points
\begin{equation}
{\mathcal R}\sup{(\lambda)}(p,P)=\frac{\varepsilon\sup{(\lambda)}\sub
i (p)}{\sqrt{w(p)}}\int\sub Pdy\sup i\exp{(-ip\cdot y)},
\end{equation}
where by eq. \eqref{eq18} can be rewritten as
\begin{equation}
S(P)=\frac 12\int\frac{d\sup 3p}{(2\pi)\sup 3}{\mathcal
R}\sup{(\lambda)}(p,P){\mathcal R^*}\sup{(\lambda)}(p,P).\label{eq21}
\end{equation}
It is easy to see that ${\mathcal R}\sup{(\lambda)}(p,P)$ is proportional to 
the tangent $T\sup k(z,P)$ fourier transformed, and that first one have 
complete information on paths in the same way that the tangent have it.

Then, we define the integration measure by means of the symbol
\begin{equation}
\int\mathcal{DR}\equiv\int\frac{\mathcal{D{R}\sup{(\lambda)}D{R^*}
\sup{(\lambda)}}}{(2\pi)\sup 3},\label{eq22}
\end{equation}
and using eqs. \eqref{eq8}, \eqref{eq21} and \eqref{eq22} it is verified
that the ground state is formally normalized
\begin{equation}
\int\mathcal{DR}\Psi\sub 0^*(P)\Psi\sub
0(P)=\int\frac{\mathcal{D{R}\sup{(\lambda)}D{R^*}
\sup{(\lambda)}}}{(2\pi)\sup 3}\exp{(-\int\frac{d\sup 3p}{(2\pi)\sup
3}{\mathcal
R}\sup{(\lambda)}{\mathcal R^*}\sup{(\lambda)})}=1.
\end{equation}

In an identical way with the Loop Representation for the Maxwell Field
\cite{DNGT}, we can observe the protagonical role played by the polarization
vectors in the measure construction.

Let us finally show the recurrence relations for  excited energy
eigenfunctions of the Hamiltonian \eqref{eq5}. Following the analogy with
the quantum mechanical harmonic oscilator we have the general solution
\begin{equation}
\Psi(P)=e\sup{-S(P)}h(P),\label{eq24}
\end{equation} 
where $h(P)$ are $P$-functional polynomials of the form
\begin{equation}
h(P)=\sum_{s=0}\int\sub Pdt\sub 1\sup{a\sub 1}\cdots\int dt\sub
s\sup{a\sub s}\Theta\sub{a\sub 1\cdots a\sub s}(t\sub 1,\ldots,t\sub
s).\label{eq25}
\end{equation}
These are the functional extension of the Hermite's polynomials, and we
demand the following properties: $\left.\Theta\sub{a\sub 1\cdots a\sub
s}\right|\sub{s=0}=\Theta\sub 0=const.$ and $\Theta\sub{a\sub 1\cdots a\sub
s}(t\sub 1,\ldots,t\sub s)$ bounded. In eq. \eqref{eq25}, the ``propagators'' 
$\Theta\sub{ab\cdots c}(t\sub a,t\sub b,\ldots,t\sub c)$
satisfy the additional symetry property
\begin{equation}
\Theta\sub{a\sub1\cdots a\sub i a\sub{i+1}\cdots a\sub s}(t\sub
1,\ldots,t\sub i,t\sub{i+1},\ldots,t\sub s)=\Theta\sub{a\sub1\cdots
a\sub{i+1}
a\sub i\cdots a\sub s}(t\sub
1,\ldots,t\sub{i+1},t\sub i,\ldots,t\sub s).
\end{equation}

In this way , using eqs. \eqref{eq9}, \eqref{eq24} and \eqref{eq25} in
$H\Psi(P)=E\Psi(P)$ we obtain the desired recurrence relations for
$s=0,1,2,\ldots$
\begin{eqnarray}
&&\big\langle\frac{(s+2)(s+1)}2\left[\Theta\sub{a\sub 1\cdots a\sub s
kl,kl}(t\sub 1,\ldots,z,z)-\Theta\sub{a\sub 1\cdots a\sub s
kk,ll}(t\sub 1,\ldots,z,z)-M\sup 2\Theta\sub{a\sub 1\cdots a\sub s kk}
(t\sub 1,\ldots,z,z)
\right]\nonumber\\
&&+2s\left[\left(\xi\sub{a\sub 1l,k}(t\sub 1-z)-\xi\sub{a\sub
1k,l}(t\sub 1-z)\right)\Theta\sub{a\sub 1\cdots a\sub s l,k}(t\sub
2,\ldots,z)+M\sup 2\xi\sub{a\sub 1k}(t\sub 1-z)\Theta\sub{a\sub 1\cdots a\sub s
k}(t\sub 2,\ldots,z)\right]\bigg\rangle\nonumber\\
&&=\pare{E-E\sub =}\Theta\sub{a\sub
1\cdots a\sub s}(t\sub 1,\ldots,t\sub s)\label{eq27}
\end{eqnarray}
with $\left.\Theta\sub{a\sub m\cdots a\sub n k\cdots l}\right|\sub{n<
m}=\Theta\sub{k\cdots l}$. Eq. \eqref{eq27} constitute a $P$-functional
generalization of the recurrence relations that generate different Hermite's
polynomials, linking the order $s+2$ functions with those of order $s$. In
this sense, the first excited state (meson state) in the Open Path
Representation depends on the first Hermite's $P$-functional polynomial,
that is to say
\begin{equation}
\Psi\sub 1(P)=\Psi\sub 0(P)\int\sub Pdt\sup k\Theta\sub k(t).\label{eq28}
\end{equation}
Using the recurrence relations \eqref{eq27} with $\Theta\sub k(t)$ bounded,
the energy of this state is $E\sub 1=E\sub 0+w(p)$.

Equation \eqref{eq28} occurs in an identical way when the creation operator,
constructed from the standard plane wave expansion acts on the vacuum state
(i.e., see something similar in a functional formulation for scalar fields
in \cite{J}). However, it would be interesting to explore a posible
construction of creation-anihilation operators in a pure
geometrical language considering $\mathcal R\sup{(\lambda)}$ as a canonical
variable (which has  the whole information about the paths), and definig some
$P$-functional derivative on this variable that acts on wave functions. The
construction of this particular derivative perhaps needs the definition of a
``$P$-functional Dirac delta'' which compares two arbitrary paths and
vanishes if these have no intersections.

I want to thank Professors L. Leal (Universidad Central de Venezuela) and C.
Di Bartolo (Universidad Sim\'on Bolivar) for many and fruitful discussions.

\end{document}